\documentclass[
  aps,
  prfluids,
  amsmath,amssymb,
  superscriptaddress,
  floatfix,
  preprint
]{revtex4-2}

\usepackage{graphicx}
\usepackage{dcolumn}
\usepackage{bm}
\usepackage{xcolor}

\begin{document}

\title{Multi-branch Shell Models of Two-Dimensional Turbulence exhibit Dual Energy–Enstrophy Cascades}

\author{Flavio Tuteri}
\email{flavio.tuteri@phys.ens.psl.eu}
\affiliation{Laboratoire de Physique de l’École normale supérieure, ENS, Université PSL, CNRS, Sorbonne Université, Université Paris Cité, Paris, F-75005, France}
\author{Sergio Chibbaro}
\affiliation{Universit\'e Paris-Saclay, CNRS, UMR 9015, LISN, F - 91405 Orsay cedex, France}

\author{Alexandros Alexakis}
\affiliation{Laboratoire de Physique de l’École normale supérieure, ENS, Université PSL, CNRS, Sorbonne Université, Université Paris Cité, Paris, F-75005, France}

\date{\today}


\begin{abstract}
Classical shell models of turbulence do not display dual cascade — inverse of energy and direct of enstrophy — because they fail to reproduce the right thermal spectra. 
We propose here a multi-branch shell model, including a geometry hierarchically organized across scales, in order to overcome this limitation. 
For this model, we demonstrate numerically both the agreement of the thermal spectra with those of two-dimensional fluid equations and the emergence of a statistically stationary dual cascade.
This construction also allows us to study local transfers and to investigate both self-similarity and non-Gaussianity. 
\end{abstract}


\maketitle


\section{\label{sec1} Introduction}

Reduced models of nonlinear chaotic systems with many degrees of freedom and interactions play a fundamental role in theoretical development~\cite{bohr1998dynamical}. 
They provide valuable laboratories for testing ideas, provided that these reduced systems are able to capture, at least qualitatively, the phenomenology of the underlying physical systems. 
For Navier–Stokes systems, shell models, based on Fourier shell averaging, have been  used to develop and test theories of turbulence \cite{mailybaev2023,aumaitre2024,biferale2003,ditlevsen2010turbulence}. 
Shell models of three-dimensional turbulence successfully reproduce the forward energy and helicity cascade
as well as the presence of intermittency. \medskip

However, the dual cascade of energy and enstrophy of two-dimensional turbulence has been difficult to reproduce. 
In two dimensions, enstrophy (the squared curl of the velocity) cascades to small scales, while energy cascades inversely to large scales, 
leading to a $k^{-3}$ spectrum for scales smaller than the forcing scale and a $k^{-5/3}$ spectrum for scales larger than the forcing scale \cite{boffetta2012two}. 
The failure of shell models to reproduce these cascade processes in two dimensions is linked to the more serious 
inability to reproduce the correct equilibrium states. Equilibrium states are states achieved in finite wavenumber systems 
in the absence of any forcing and dissipation that lead under certain conditions to equipartition of 
conserved quantities among all degrees of freedom. In the (truncated) two dimensional Navier-Stokes equation 
the equilibrium states lead to a $k^1$ spectrum for the equipartition of energy and a $k^{-1}$ for the equipartition of enstrophy
\cite{kraichnan1975statistical,kraichnan1980two,alexakis2018}.
Classical shell models however, result in an equivalent $k^{-1}$ and a $k^{-3}$ spectrum for the equipartition of energy and enstrophy 
respectively. The fact that the enstrophy equipartition spectrum coincides with the forward cascade of enstrophy spectrum 
prevents the dual cascade of energy and enstrophy from manifesting itself \cite{gilbert2002inverse}. 
As an alternative, to achieve an inverse cascade in shell models  that mimics the dual cascade scenario of two-dimensional turbulence, it is imposed to conserve a pseudo-enstrophy that has physical dimensions of velocity squared times $k^{2\alpha}$, with $\alpha<1$ \cite{boffetta2005,chen2024odd,benavides2025phase}.  
This alternative does indeed reproduce the dual cascade scenario, but with spectra that do not coincide with the two dimensional turbulence predictions.
\medskip

This difficulty can be overcome 
for a new class of models with a hierarchical organization of spatial degrees of freedom across scales.
Such shell models were introduced long ago \cite{aurell1994,aurell1997,benzi1997} and have recently been reexamined \cite{ajzner2023,tuteri2026}. 
These models predict the correct equilibrium spectra, and 
the direct cascade of enstrophy is possible as shown in \cite{aurell1997}.
\medskip

In this work we show 
that these models can reproduce the dual cascade scenario
with a forward cascade of enstrophy and an inverse cascade of energy.




\section{\label{sec2} Model}

We recall that a shell model is heuristically a reduction of the degrees of freedom of the Navier–Stokes equations, obtained through a spherical coarse-graining in Fourier space that retains only quasi-local triadic interactions \cite{yamada1987,biferale2003}. 
Nonetheless, geometry plays a fundamental role in turbulence, and the introduction of additional spatial degrees of freedom has been previously proposed \cite{aurell1994,aurell1997,benzi1997,ajzner2023,tuteri2026}.
Here we assume a hierarchical organization of space across scales.


\subsection{\label{subsec2.1} Definition}
\begin{figure*}
\centering
\includegraphics[width=\textwidth]{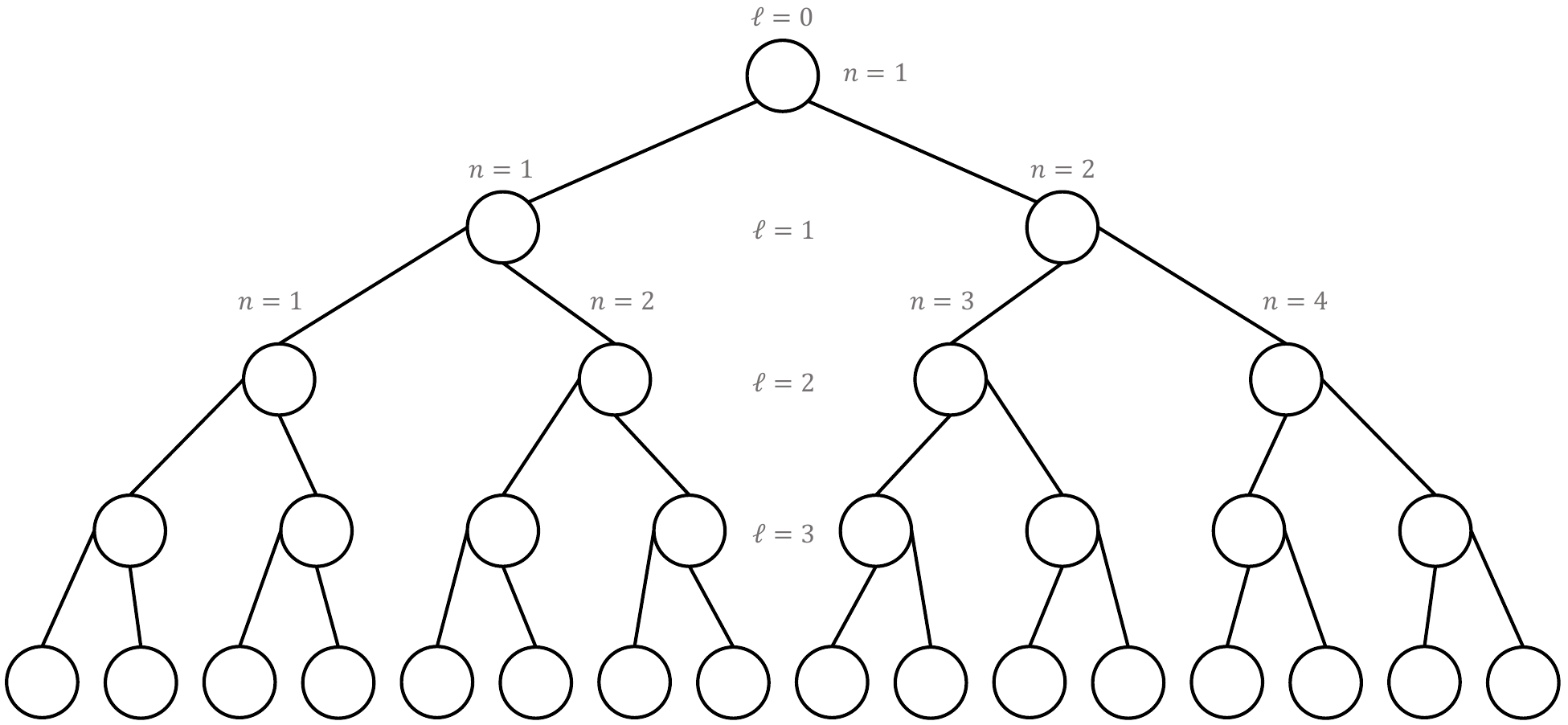}
\caption{\label{figTree}
Schematic representation of the dyadic shell model topology.
Each level $\ell$ corresponds to a shell with characteristic wavenumber $k_\ell=\lambda^\ell$, while the index $n$ labels spatial substructures of the same characteristic length scale.}
\end{figure*}
We introduce a shell-model framework defined on a $p$-adic tree topology.
The state is specified by complex dynamical variables $u_{\ell,n}$ associated with the nodes
\begin{equation}
(\ell,n)\quad\text{with}\quad
\ell\in\{0,\ldots,L\},\;
n\in\{1,\ldots,p^\ell\}.
\end{equation}
See Fig.~\ref{figTree} for the case $p=2$.
The level index $\ell$ plays the role of the scale index and is associated to the wavenumber $k_\ell=\lambda^\ell$, with inter-shell ratio $\lambda>1$.
For fixed $\ell$, the index $n$ labels spatial substructures at the corresponding characteristic length.
With this interpretation, global observables are defined as volume-weighted sums of local densities.
In particular, the total energy and enstrophy read
\begin{equation}
    E=\frac{1}{2}\sum_{\ell=0}^{L} {k_\ell}^{-D}\sum_{n=1}^{p^\ell}\lvert u_{\ell,n}\rvert^2,\qquad
    Z=\frac{1}{2}\sum_{\ell=0}^{L} {k_\ell}^{2-D}\sum_{n=1}^{p^\ell}\lvert u_{\ell,n}\rvert^2,
\end{equation}
where $D$ denotes the effective spatial dimension. Note that the factor ${k_\ell}^{-D}$ expresses the $D$-dimensional volume 
of the structure at scale $\ell$. Accordingly, the energy spectrum $E_\ell$ is defined as
\begin{equation}
    E_\ell =\frac{1}{2} {k_\ell}^{-D}\sum_{n=1}^{p^\ell}\lvert u_{\ell,n}\rvert^2.
\end{equation}

Interactions between nodes have to be compatible with genealogy. Each node $(\ell,n)$, except the root $(0,1)$, has a single parent $(\ell-1,\lceil n/p\rceil)$, where $\lceil x \rceil =\min\{m\in\mathbb{Z}\mid m\ge x\}$ is the ceiling. Each node $(\ell,n)$, except the leaves $(L,n)$, has $p$ children, namely $(\ell+1,p(n-1)+1),\ldots,(\ell+1,pn)$.
The dynamics on this topology is defined by
\begin{equation}
\label{maineq}
    \bigg[\frac{d}{dt}+\nu{k_\ell}^{2n_\nu}+\alpha {k_\ell}^{-2n_\alpha}\bigg]u_{\ell,n}=i N_{\ell,n}[u]+f_{\ell,n},
\end{equation}
where $\nu$ and $n_\nu$ fix the hyper-viscous dissipation, $\alpha$ and $n_\alpha$ define the hypo-viscous drag, $f_{\ell,n}$ is the forcing, and $N_{\ell,n}$ the nonlinear coupling. 
Our model generalizes the Sabra coupling~\cite{lvov1998} by equally distributing admissible triads along genealogical branches:
\begin{alignat}{3}
    N_{\ell,n}=k_\ell\bigg[&a\,\lambda\,\frac{1}{p}\sum_{h=0}^{p-1}\bigg( &&{u_{\ell+1,pn-h}              }^\ast \,\frac{1}{p}\sum_{m=0}^{p-1} &&u_{\ell+2,p(pn-h)-m}\bigg)\nonumber\\
                  +\,&b                                             &&{u_{\ell-1,\lceil n/p  \rceil}}^\ast \,\frac{1}{p}\sum_{m=0}^{p-1} &&u_{\ell+1,pn-m}            \nonumber\\
                  -\,&c\,\lambda^{-1}                               && u_{\ell-2,\lceil n/p^2\rceil}                               &&u_{\ell-1,\lceil n/p\rceil}\bigg],
\label{nonlinear}
\end{alignat}
supplemented with the boundary conditions $u_{-2,\bullet}=u_{-1,\bullet}=0=u_{L+1,\bullet}=u_{L+2,\bullet}$.
Homogeneous solutions, i.e. constant in $n$, correspond to the classical single-branch model.
The inviscid and unforced dynamics conserves the total energy provided that
\begin{equation}
    a+b+c=0,
\label{energy_conservation}
\end{equation}
together with the condition
\begin{equation}
    D^{-1}=\log_p\lambda.
\label{spaDim}
\end{equation}
Relation~(\ref{spaDim}) expresses the geometric consistency of the cascade: the branching number $p$ equals the volume contraction between successive scales.
Requiring, in addition, the conservation of enstrophy leads to the further constraint
\begin{equation}
    a+\lambda^2 b+\lambda^4 c=0.
\label{model_parameters}
\end{equation}
The local energy transfer from $(\ell-1,\lceil n/p\rceil)$ to $(\ell,n)$ is defined as the rate of change of the energy contained in $(\ell,n)$ and in all descendant nodes:
\begin{align}
{\Pi^e}_{\ell,n}=&\;\operatorname{Im}\!\bigg[\sum_{j=\ell}^{L}{k_j}^{-D}\sum_{m=p^{j-\ell}(n-1)+1}^{p^{j-\ell}n} u_{j,m}N_{j,m}[u]^\ast\bigg] \\
             ={k_{\ell}}^{1-D}&\;\operatorname{Im}\!\bigg[-a\,u_{\ell-1,\lceil n/p\rceil}u_{\ell,n}\,\frac{1}{p}\sum_{m=0}^{p-1}{u_{\ell+1,pn-m}}^\ast \nonumber\\
             &+c\,\lambda^{-1}\, u_{\ell-2,\lceil n/p^2\rceil}u_{\ell-1,\lceil n/p\rceil}{u_{\ell,n}}^\ast\bigg].
\label{local_flux_erg}
\end{align}
Expression~(\ref{local_flux_erg}) is a direct consequence of energy conservation~(\ref{energy_conservation}).
Analogously, we have derived the local enstrophy flux:
\begin{align}
{\Pi^z}_{\ell,n}=&\;\operatorname{Im}\!\bigg[\sum_{j=\ell}^{L}{k_j}^{2-D}\sum_{m=p^{j-\ell}(n-1)+1}^{p^{j-\ell}n} u_{j,m}N_{j,m}[u]^\ast\bigg] \\
             ={k_\ell}^{3-D}&\;\operatorname{Im}\!\bigg[-\lambda^{-2}a\,u_{\ell-1,\lceil n/p\rceil}u_{\ell,n}\,\frac{1}{p}\sum_{m=0}^{p-1}{u_{\ell+1,pn-m}}^\ast \nonumber\\
             &+c\,\lambda^{-1}\, u_{\ell-2,\lceil n/p^2\rceil}u_{\ell-1,\lceil n/p\rceil}{u_{\ell,n}}^\ast\bigg].
\label{local_flux_ens}
\end{align}
In a statistically stationary regime, the system may sustain a dual cascade characterized by constant mean fluxes across scales,
\begin{equation}
    \langle {\Pi^e}_{\ell,n}\rangle_{n,t}=\epsilon_e \quad \text{for } k_\ell \ll k_f,
    \qquad
    \langle {\Pi^z}_{\ell,n}\rangle_{n,t}=\epsilon_z \quad \text{for } k_\ell \gg k_f,
\end{equation}
where $k_f$ denotes the forcing wavenumber, while $\epsilon_e$ and $\epsilon_z$ are the mean energy and enstrophy injection rates, respectively. 


\subsection{\label{subsec2.2} Equilibrium}

\begin{figure*}
\centering
\includegraphics[width=\textwidth]{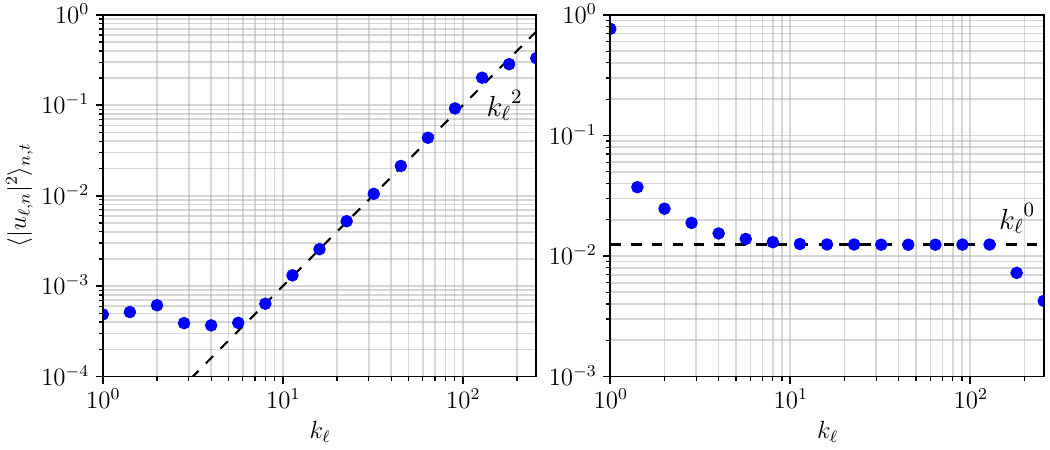}
\caption{\label{figThermal}
Thermal spectra obtained from the inviscid and unforced dynamics for $D=2$. 
At equilibrium, the scaling (\ref{eqScaling}) is recovered. 
Left: initial condition with energy concentrated at small scales. 
Right: initial condition with energy concentrated at large scales.
}
\end{figure*}

To assess the possibility of a dual cascade, direct for enstrophy and inverse for energy, we analyze the corresponding Gibbs equilibrium states.
In our interpretation, when a thermal state exists with the same spectral scaling as a cascade solution, the nonlinear dynamics can satisfy the spectral constraint through local thermalization, without sustaining a persistent inter-shell transfer.
As a result, the mean flux across scales vanishes.
For our model, the inviscid and unforced dynamics conserves two quadratic quantities, so that the corresponding Gibbs equilibrium measure is
\begin{equation}
    P[u]\propto \exp[-\beta E - \gamma Z] =\exp\bigg\{-\frac{1}{2}\sum_{\ell=0}^L 
    \bigg[ \Big(\beta{k_\ell}^{-D}+\gamma k_\ell^{2-D}\Big)
    \sum_{n=1}^{p^\ell}\lvert u_{\ell,n}\rvert^2\bigg]\bigg\},
\end{equation}
with $\beta,\gamma>0$.
Under this measure,
\begin{equation}
\label{eqScaling}
    E_\ell =\frac{1}{k_\ell^D} \sum_n \langle\lvert u_{\ell,n}\rvert^2\rangle_P =
    \langle\lvert u_{\ell,n}\rvert^2\rangle_P
    =\frac{{k_\ell}^{D}}{\beta+\gamma k_\ell^{2}},
\end{equation}
where in the second equality we have used the fact that the mode energy is independent of $n$, together with equation (\ref{spaDim}).
Extreme cases for thermal spectrum are:
\begin{equation}
\label{eqSpc1}
    E_\ell \propto {k_\ell}^{D},
\end{equation}
corresponding to equipartition of energy, and
\begin{equation}
\label{eqSpc2}
    E_\ell \propto {k_\ell}^{D-2},
\end{equation}
corresponding to equipartition of enstrophy.
%
For $D=2$, the present model yields thermal spectra $E_\ell \propto {k_\ell}^{-2}$ and $E_\ell \propto {k_\ell}^{-0}$ 
consistent with those of two-dimensional Navier--Stokes, 
once the shell-averaging factor is properly taken into account. In Figure \ref{figThermal} we have numerically verified prediction~(\ref{eqSpc1})--(\ref{eqSpc2}) for the $2$-adic model with $D=2$.
In these runs, simulations of model (\ref{maineq}) were performed with no forcing and no dissipation 
and initial conditions that had most of the energy in the small wavenumbers (left panel) 
and most of the energy in the large wavenumbers (right panel).
At late times the thermal energy spectra were reproduced. 

\section{Results}

\begin{figure*}
\centering
\includegraphics[width=\textwidth]{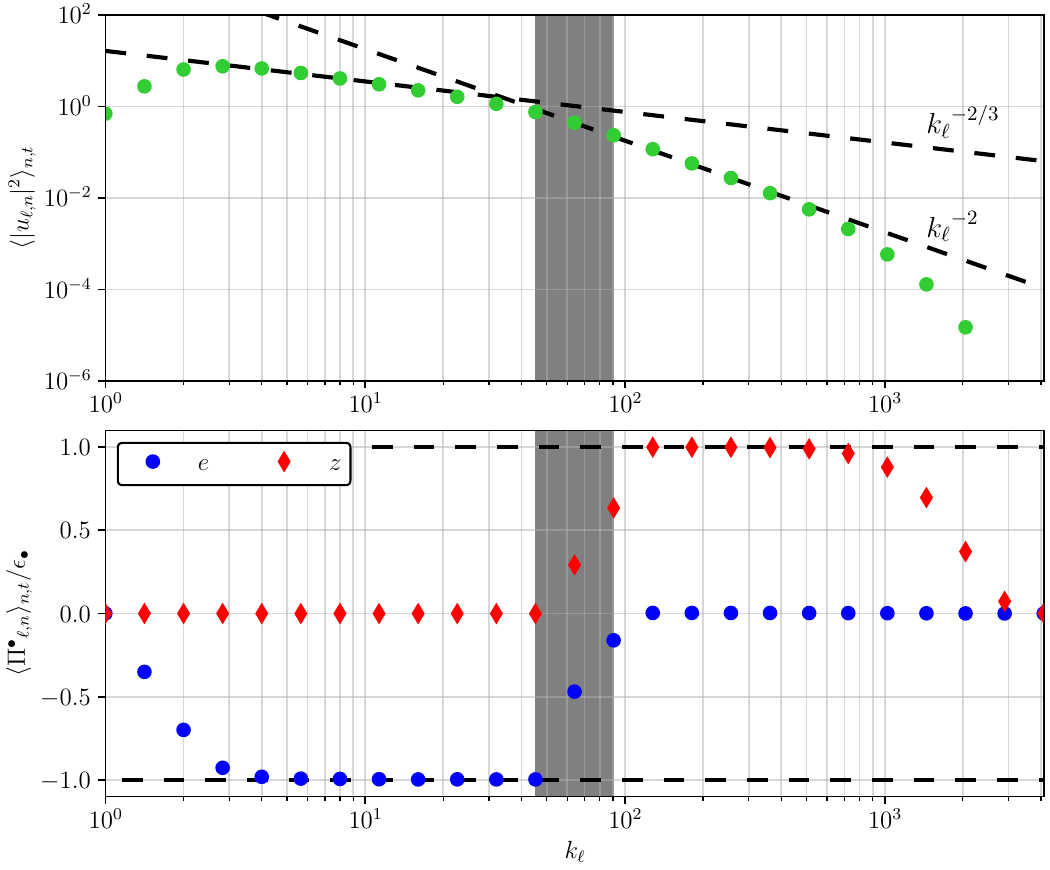}
\caption{\label{figSpectra}
Top: mean energy spectrum. The forcing band is shaded in grey. 
At scales larger than the forcing scale, a $k_\ell^{-2/3}$ scaling consistent with an inverse energy cascade is observed, while at smaller scales the spectrum approaches a $k_\ell^{-2}$ scaling consistent with a direct enstrophy cascade. 
Bottom: mean fluxes of the quadratic invariants, energy (red) and enstrophy (blue). 
}
\end{figure*}

We report the results of numerical simulations evolving (\ref{maineq}) with $2$-adic branching,
with intershell ratio $\lambda=\sqrt{2}$ ($\Leftrightarrow D=2$), 
hyperviscosity $\nu=10^{-12}$ with $n_\nu=2$, 
large-scale drag $\alpha=0.5$ with $n_\alpha=2$, 
and forcing injecting energy in shells $\ell_f=11$–$13$ over a total of $25$ shells.
Figure~\ref{figSpectra} shows the mean energy spectrum (top) and the corresponding mean fluxes of energy and enstrophy (bottom). 
The spectrum exhibits a ${k_\ell}^{-2/3}$ scaling at scales larger than the forcing band and a ${k_\ell}^{-2}$ scaling at smaller scales. 
These scalings are consistent with an inverse energy cascade and a direct enstrophy cascade, respectively, as confirmed by the behavior of the mean fluxes in the lower panel. \medskip

\begin{figure*}
\centering
\includegraphics[width=\textwidth]{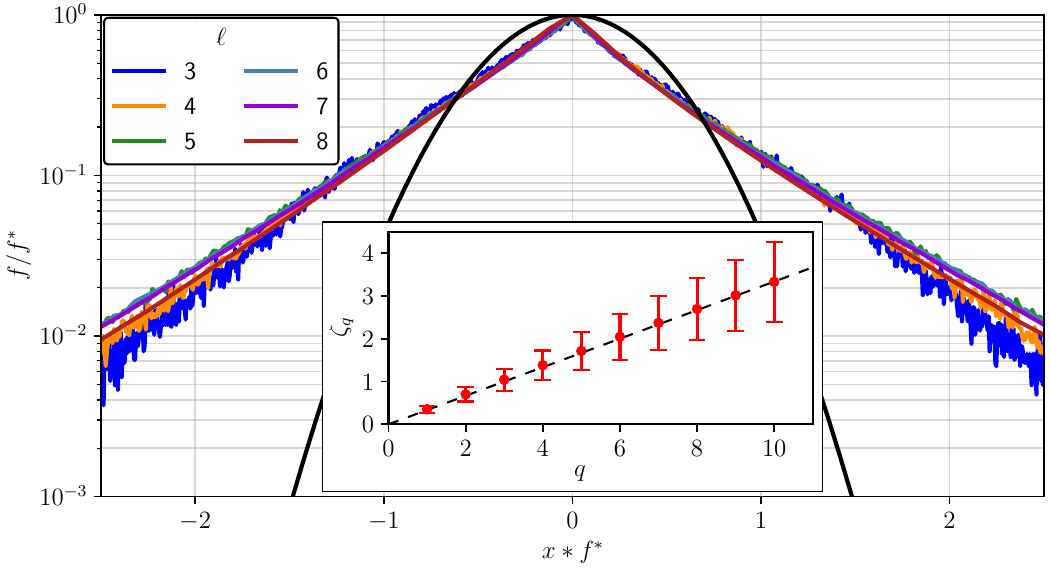}
\caption{\label{figSim} 
Probability density functions $f$ of the local energy flux ${\Pi^e}_{\ell,n}$ for $\ell$ within the inverse inertial range, rescaled by their maxima $f^*$. The Gaussian reference is shown in black.
Inset: scaling exponents computed from the local energy flux structure functions; the dashed line indicates the Kolmogorov dimensional prediction.
}
\end{figure*}

To investigate the inverse cascade inertial range, we analyze the statistics of the local energy flux. 
Figure~\ref{figSim} shows the probability density functions of the local energy flux ${\Pi^e}_{\ell,n}$ at different scales $\ell$. 
Assuming spatial ergodicity, different values of $n$ are interpreted as independent samplings of the same random variable. 
Because the number of nodes at level $\ell$ scales as $2^\ell$, smaller values of $\ell$ provide fewer statistical samples. The higher noise observed for PDFs at small $\ell$ is therefore a finite-sample effect rather than a dynamical feature.
The distributions are rescaled by their maxima, and the collapse is observed, indicating self-similarity despite strongly non-Gaussian statistics, as observed for the local energy flux in direct numerical simulations of two-dimensional Navier-Stokes turbulence \cite{xiao2009,boffetta2000inverse}. 
The inset shows the scaling exponents of the local energy-flux structure functions, computed following~\cite{deWit2024}; the measured exponents follow the dimensional prediction, based on self-similarity.

\section{Conclusions}

We have studied a multi-branch shell model with a hierarchical organization of spatial degrees of freedom across scales. 
Within this class of models, dimensional analysis allows the identification of a configuration whose thermal equilibrium reproduces the scaling expected for two-dimensional fluids. \medskip

We confirm that the equilibrium predictions hold for this model
and we demonstrate for the first time in a shell model the emergence of a statistically stationary turbulent regime characterized by a dual cascade: an inverse cascade of energy toward large scales and a direct cascade of enstrophy toward small scales. 
The corresponding inertial ranges are clearly identified through the behavior of the energy spectrum and the fluxes of the two quadratic invariants. \medskip

The present setup can then be applied to model quasi-two dimensional flows as in \cite{boffetta2012two,chen2024odd} 
keeping however the correct dimensions of enstrophy and the correct equilibrium spectra.
Another key advantage of the present framework is the possibility of defining local transfers of conserved quantities along the hierarchical structure. 
This allows a detailed statistical analysis of the cascade process. 
In particular, we observe self-similar behavior of the local energy flux together with strongly non-Gaussian fluctuations, consistent with observations in direct numerical simulations of two-dimensional Navier–Stokes turbulence. \medskip

The examined model therefore provides a minimal framework in which geometry, equilibrium properties, and cascade dynamics are consistent with the archetypal phenomenology, offering a promising tool for the theoretical investigation of systems displaying an inverse energy cascade, such as geophysical fluids.


\begin{acknowledgments}
This work received financial support from the CNRS through the MITI interdisciplinary initiatives, under its exploratory research program. SC acknowledges funding from the ANR grant SCALP  (ANR-24-CE23-1320).
\end{acknowledgments}



\bibliography{apssamp}

\end{document}